\documentclass[letter,longauth,traditabstract]{aa}
\newcommand{\as}{$^{\prime\prime}$}
\newcommand{\am}{$^{\prime}$}
\newcommand{\um}{$\mu$m}
\newcommand{\cmg}{cm$^{2}$ g$^{-1}$}
\newcommand{\msun}{M$_{\odot}$}

\usepackage{epsfig,graphicx,color}
\usepackage{txfonts}
\begin{document}

\title{Small-scale structure in the Rosette molecular cloud \\
revealed by {\it Herschel}\thanks{{\it Herschel}\/ is an ESA space observatory with science instruments provided by European-led Principal Investigator consortia and with important participation from NASA.}}

\author{J. Di Francesco\inst{1,2},
          S. Sadavoy\inst{2,1},
          F. Motte\inst{3},
          N. Schneider\inst{3},
          M. Hennemann\inst{3},
          T. Csengeri\inst{3},
          S. Bontemps\inst{3,4},
          Z. Balog\inst{5},
          A. Zavagno\inst{6},
          Ph. Andr\'e\inst{3},
          P. Saraceno\inst{7},
          M. Griffin\inst{8},
          A. Men'shchikov\inst{3},
          A. Abergel\inst{9},
          J.-P. Baluteau\inst{6},
          J.-Ph. Bernard\inst{10},
          P. Cox\inst{11},
          L. Deharveng\inst{6},
          P. Didelon\inst{3},
          A.-M. di Giorgio\inst{7},
          P. Hargrave\inst{8},
          M. Huang\inst{12},
          J. Kirk\inst{8},
          S. Leeks\inst{13},
          J. Z. Li\inst{12},
          A. Marston\inst{14},
          P. Martin\inst{15},
          V. Minier\inst{3},
          S. Molinari\inst{7},
          G. Olofsson\inst{16},
          P. Persi\inst{17},
          S. Pezzuto\inst{7},
          D. Russeil\inst{6},
          M. Sauvage\inst{3},
          B. Sibthorpe\inst{18},
          L. Spinoglio\inst{7},
          L. Testi\inst{19},
          D. Teyssier\inst{14},
          R. Vavrek\inst{14},
          D. Ward-Thompson\inst{8},
          G. White\inst{13,20},
          C. Wilson\inst{21}, and
          A. Woodcraft\inst{18}}

   \institute{
            National Research Council of Canada, Herzberg Institute of Astrophysics, 5071 West Saanich Rd., Victoria, BC, V9E 2E7, Canada\\
            \email{james.difrancesco@nrc-cnrc.gc.ca}
            \and 
            University of Victoria, Department of Physics and Astronomy, PO Box 3055, STN CSC, Victoria, BC, V8W 3P6, Canada
            \and
Laboratoire AIM, CEA/IRFU -- CNRS/INSU -- Universit\'e Paris Diderot, CEA-Saclay, F-91191 Gif-sur-Yvette Cedex, France
            \and
             Laboratoire d'Astrophysique de Bordeaux, CNRS/INSU -- Universit\'e de Bordeaux, BP 89, 33271 Floirac cedex, France
            \and
             Max-Planck-Institut f\"ur Astronomie, K\"onigstuhl 17, D-69117, Heidelberg, Germany
            \and
             Laboratoire d'Astrophysique de Marseille, CNRS/INSU - Universit\'e de Provence, 13388 Marseille cedex 13, France
            \and
             INAF-IFSI, Fosso del Cavaliere 100, 00133 Roma, Italy
            \and
              Cardiff University School of Physics and Astronomy, UK
            \and
             IAS, Universit\'e Paris-Sud, 91435 Orsay, France
            \and
             CESR \& UMR 5187 du CNRS/Universit\'e de Toulouse, BP 4346, 31028 Toulouse Cedex 4, France
            \and
             IRAM, 300 rue de la Piscine, Domaine Universitaire, 38406 Saint Martin d'H\'eres, France
            \and
             National Astronomical Observatories, Chinese Academy of Sciences, Beijing 100012, China
            \and
             The Rutherford Appleton Laboratory, Chilton, Didcot, Oxon OX11 ONL, UK
            \and
             Herschel Science Centre, ESAC, ESA, PO Box 78, Villanueva de la Ca\~nada, 28691 Madrid, Spain
            \and
             CITA \& Department of Astronomy and Astrophysics, University of Toronto, 50 St. George St., Room 101, Toronto, ON, M5S 3H4, Canada
            \and
             Department of Astronomy, Stockholm Observatory, AlbaNova University Center, Roslagstullsbacken 21, 10691 Stockholm, Sweden
            \and
             INAF-IASF, Sez. di Roma, via Fosso del Cavaliere 100, 00133, Rome, Italy
            \and
             UK Astronomy Technology Centre, Royal Observatory Edinburgh, Blackford Hill, EH9 3HJ, UK
            \and
             ESO, Karl Schwarzchild Str. 2, 85748, Garching, Germany
            \and
             Department of Physics \& Astronomy, The Open University, Milton Keynes MK7 6AA, UK
            \and
             Department of Physics and Astronomy, ABB-241, McMaster University, 1280 Main St. W., Hamilton, ON, L8S 4M1, Canada}

\offprints{J. Di Francesco}

\mail{james.difrancesco@nrc-cnrc.gc.ca}

\titlerunning{Small-scale structure in the RMC} 

\authorrunning{J.~Di~Francesco et al.}

\date{Received 2010 March 31 ; accepted 2010 May 7}

\abstract
{We present a preliminary analysis of the small-scale structure found in new 
70-520 \um\ continuum maps of the Rosette molecular cloud (RMC), obtained 
with the SPIRE and PACS instruments of the {\it Herschel} Space Observatory.  
We find 473 clumps within the RMC using a new structure identification 
algorithm, with sizes up to $\sim$1.0 pc in diameter.  A comparison with 
recent {\it Spitzer} maps reveals that 371 clumps are ``starless" (without 
an associated young stellar object), while 102 are ``protostellar." Using 
the respective values of dust temperature, we determine the clumps have 
masses ($M_{C}$) over the range -0.75 $\leq$ log~($M_{C}$/\msun) $\leq$ 2.50.  
Linear fits to the high-mass tails of the resulting clump mass spectra (CMS) 
have slopes that are consistent with those found for high-mass clumps 
identified in CO emission by other groups.}  


\maketitle

\section{Introduction} \label{intro}

Stars form within molecular clouds, after some fraction of cloud material is 
first condensed into smaller-scale structures.  How this process unfolds is 
not well understood, however.  For example, what conditions within clouds drive 
the formation of small-scale structures that in turn produce stars of any given 
mass, e.g., high-mass stars.  Detailed observations of the small-scale 
structure within molecular clouds should provide valuable insight into how 
stars of various masses form out of dense material. 

The need for sensitive probes of small-scale structure in molecular clouds 
is the impetus behind the ``{\it Herschel}\/ OB Young Stellar objects" (HOBYS) 
Key Project (see Motte et al. 2010).  The HOBYS team is currently using the ESA
{\it Herschel} Space Observatory (Pilbratt et al. 2010) to obtain wide-field 
maps of $\sim$15 molecular clouds which are forming high-mass stars within 3 
kpc of the Sun.  These clouds are being observed with both the {\it Herschel} 
SPIRE (Griffin et al.\/ 2010) and PACS (Poglitsch et al.\/ 2010) instruments, 
ultimately to obtain wide maps of these clouds at 70-520 \um\ at 
diffraction-limited resolutions of $\sim$18\as\ $\times$ ($\lambda$/250 \um).  
Such data sample thermal emission from cold dust grains mixed with 
molecular gas within the cloud.  Since continuum emission at submillimetre 
wavelengths is typically optically thin, detections of continuum emission 
from dust at high resolution can trace the organization of mass in clouds 
on small scales.  By sensitively sampling emission across the peaks of the 
spectral energy distributions (SEDs) of this dust, SPIRE and PACS data can 
constrain both the column density and temperature of the dust and provide 
unparalleled censuses of the small-scale structure in molecular clouds. 

In this paper, we describe the small-scale structure detected with SPIRE and
PACS in the Rosette molecular cloud (RMC), the first of the HOBYS target list 
that was observed as part of the early {\it Herschel} ``science demonstration 
phase" (SDP) campaign.  The RMC is a rich location of star formation at a 
distance of 1.6 kpc (Johnson 1962; P\'erez et al. 1987; Park \& Sung 2002).
The cloud is southeast of NGC 2244, the Rosette Nebula, and the expanding 
HII region associated with NGC 2244 has begun to interact with it.  The 
structure of material within the RMC has been previously examined by Williams, 
Blitz \& Stark (1995), Schneider et al.\/ (1998) and Dent et al.\/ (2009) 
using various CO observations.  Alternatively, the populations of young 
stellar objects (YSOs) within the RMC have been studied by Phelps \& Lada 
(1998), Li \& Smith (2005), Poulton et al.\/ (2008), and Rom\'an-Z\'u\~niga 
et al.\/ (2008) using infrared observations.  Notably, these studies identified 
10 embedded clusters within the RMC, i.e., PL01-07 and REFL08-10.  A recent 
review of observations of the RMC has been made by Rom\'an-Z\'u\~niga \& Lada 
(2008; RZL).  

The SPIRE and PACS observations described in this paper provide a preliminary
look into how {\it Herschel}\/ data trace small-scale structures in the RMC, 
and the relationships between these structures and YSOs.  Analyses of other 
aspects within the SDP data of the RMC can be found in the accompanying papers 
by Motte et al.\/ 2010, Henneman et al.\/ 2010, and Schneider et al.\/ 2010.)

\section{Observations} \label{obs}

A complete description of the observations described here can be found in
the accompanying paper by Motte et al. (2010).  In short, an extinction map 
of the RMC was first constructed for planning purposes using 2MASS data and 
the AvMAP algorithm.  AvMAP derives extinction maps with typical resolutions 
of $\sim$2\am\ using colour excesses in a manner akin to the NICER 
algorithm of Lombardi \& Alves (2001).  AvMAP, however, also uses a stellar 
model to predict foreground source densities; see Schneider et al. (2010a).
Using the AvMAP data as a guide, an area of the RMC $\approx$ 1 deg$^{2}$ in 
extent encompassing areas southeast of NGC 2244 with $A_{V}$ $>$ 5 was mapped 
in common with SPIRE and PACS in parallel scan mode on 2009 October 20.  Maps 
were obtained simultaneously at 70 $\mu$m and 170 $\mu$m with PACS and at 
250 $\mu$m, 350 $\mu$m, and 520 $\mu$m with SPIRE.  

\onlfig{1}{
\begin{figure*}
\centering
\includegraphics[width=17cm]{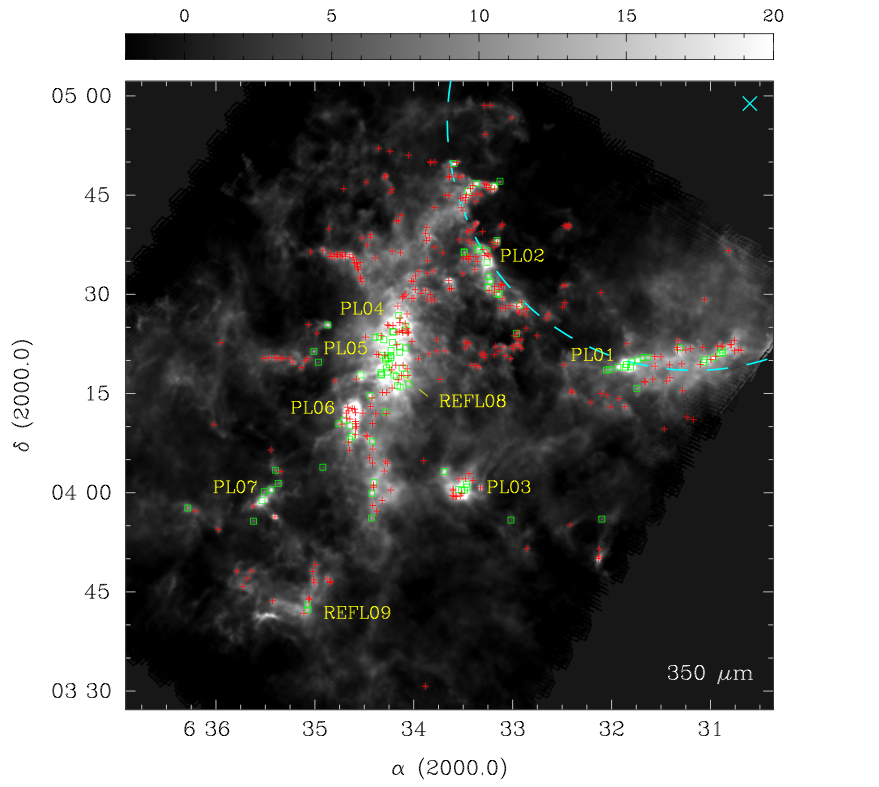}
\caption [] {{\it Herschel} map of the Rosette molecular cloud at 350 \um.  
The positions of 473 clumps identified by {\it getsources}\/ are shown, where 
red crosses denote starless clumps and green boxes denote protostellar clumps.  
The cyan dashed line indicates the approximate edge of the Rosette Nebula 
and the cyan cross indicates the central position of NGC 2244 from Table 3 of 
RZL.  The nine embedded clusters identified by Phelps \& Lada (1998; PL) and 
Rom\'an-Z\'u\~niga et al. (2008; REFL) in the {\it Herschel} field are also
labelled.} 
\label{rosette-map}
\end{figure*}
}

The SPIRE and PACS data were reduced separately using common routines of the 
{\it Herschel}\/ Interactive Processing Environment\footnote{HIPE is a joint 
development by the Herschel Science Ground Segment Consortium, consisting of
ESA, the NASA Herschel Science Center, and the HIFI, PACS and SPIRE consortia.}
(HIPE; see Ott 2010).  To facilitate the identification and analysis of the 
cold, small-scale structure in the RMC, i.e., dense clumps, the PACS maps were 
Gaussian smoothed to the 18\as\ FWHM resolution of the 250 \um\ SPIRE map.  
A full analysis of the unsmoothed PACS maps will be provided in the near future 
by Sadavoy et al.\/ (2010b, in preparation).

\section{Results} \label{results}

Figure 1 (available electronically) shows the 350 $\mu$m map of the RMC, 
revealing a rich amount of structure in continuum emission.  (See Motte et 
al.\/ for maps at all five wavelengths.)  Several bright knots are seen 
embedded in large filaments, and each can be associated with known infrared 
clusters (see Figure 1, also RZL).  Ubiquitous, faint diffuse emission is 
also seen, and in many cases this appears filamentary as well.  Of special 
note in Figure 1 is the filamentary arc seen in the northwestern corner, 
indicating the ``ridge" of the RMC (+ ``Core D"), i.e., where the adjacent 
NGC 2244 HII region (the Rosette Nebula) is interacting with the molecular 
cloud.  

From the (smoothed) PACS and SPIRE maps, 473 continuum objects were identified 
using the {\it getsources}\/ algorithm; an accompanying paper in this volume 
by Men'shchikov et al.\/ (2010) describes {\it getsources} in more detail.  In 
brief, {\it getsources}\/ first decomposes maps into images over a range of 
scales and objects are identified as discrete emission regions in each image.  
Objects are then defined using information on many scales and at many 
wavelengths.  For example, positions and extents (i.e., ``footprints") are 
determined from the first and second moments of intensities in images where 
all scales are recombined.  Integrated flux densities are calculated by summing 
the background-subtracted emission under the footprint of an object.  Sizes 
are well-approximated by 2D Gaussian fits.

For simplicity, we choose to identify all objects found by {\it 
getsources}\/ as ``clumps"\footnote{The term ``clump" has been used 
previously to describe sites of grouped or clustered star formation, as 
opposed to sites of single- (or multiple-) star formation, i.e., ``cores"; 
see Williams, Blitz \& McKee (2000).  We use ``clump" here to describe all 
discrete condensations of material of size $<$ 1 pc within the larger 
cloud.}  Figure 1 shows the locations of all 473 clumps identified in 
the RMC overlaid onto the 350 \um\ map.  Most are found within the brighter 
filaments of the RMC, with neighbouring clumps tracing the linear structures 
of underlying filaments.  Several ``isolated" clumps (i.e., those without 
close neighbours), however, are also seen associated with fainter emission.  
Relatively few clumps are seen inside the northwestern ``ridge" and 
coincident with the NGC 2244 HII region.

Figure 1 also shows which clumps in the RMC contain YSOs (``protostellar 
clumps") and which do not (``starless clumps").  These determinations were 
made using {\it Spitzer} IRAC and MIPS data of the region (see Poulton et 
al.\/ 2008; Balog et al.\/ 2010, in preparation).  The criteria used to 
determine which clumps contain YSOs are based on those developed recently by 
Sadavoy et al.\/ (2010a) for small-scale structure detected at 850 $\mu$m in 
the less-distant Orion, Taurus, Perseus and Ophiuchus molecular clouds.  For 
this comparison, potential YSOs had to be detected in all four IRAC bands or 
at 24 \um.  Their colours and brightness were checked against colour criteria 
to distinguish them as either YSOs or other objects (e.g., background galaxies 
or stars).  In addition, their positions relative to clumps were checked for 
coincidence within areas defined by the 2D Gaussian fits to the clump extents.  
We do not distinguish here between protostellar Classes of YSOs in clumps; 
see Henneman et al. for further discussion.

\begin{figure}[ht]
\includegraphics[angle=0,width=90mm]{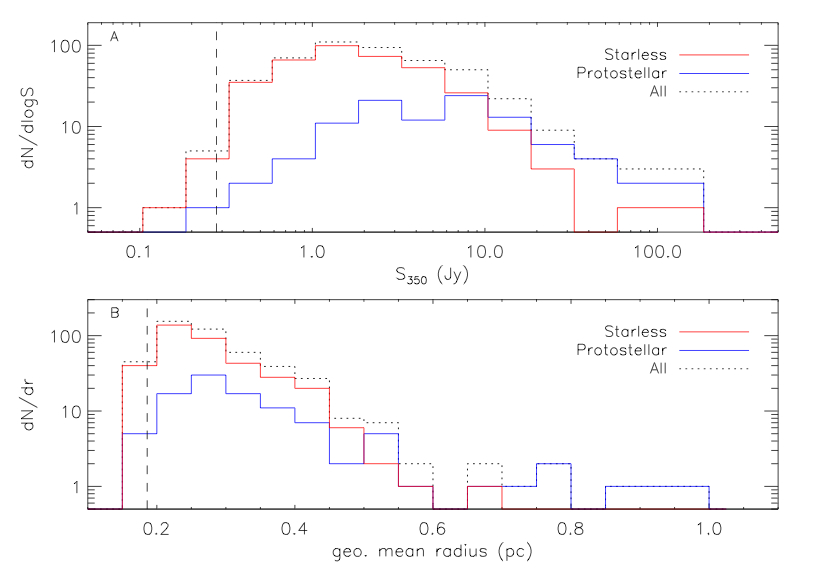}
\caption [] {Distributions of 350 \um\ fluxes (A) and geometric mean sizes (B) 
of RMC clumps.  Dotted black, solid red, and solid blue lines indicate all 
clumps, starless clumps, and protostellar clumps, respectively.  In panel A, 
the vertical dashed line indicates roughly a 3 $\sigma$ rms limit for detected 
clumps.  In panel B, the vertical dashed line indicates the resolution limit 
of the smoothed 250 $\mu$m data.}
\label{clump-dist}
\end{figure}

In total, we find 371 starless clumps and 102 protostellar clumps in the RMC.
Most protostellar clumps contain one YSO, but $\sim$35\% contain more than one.
Figure 1 suggests that brighter clumps tend to be protostellar, while fainter 
clumps tend to be starless.  (Exceptions to both cases are seen, however.)  As 
expected, the protostellar clumps are associated with the embedded clusters 
known in the cloud, i.e., PL01-07 and REFL08-10.  Many starless clumps appear 
to be associated with the RMC ridge and the area between it and the RMC 
``central core" (i.e., the concentration of bright protostellar clumps 
associated with the embedded clusters PL04-06 and REFL08 in the centre of 
Figure 1).  Relatively few ($\sim$15) clumps are seen behind the ridge itself,
and they all appear to be starless.  Excluding areas inside and outside the 
ridge, the starless and protostellar clumps appear well-mixed, suggesting
no ``age gradient" across the RMC.  (Schneider et al. 2010b suggest, however, 
that the most massive cores in the RMC decrease in age with increased distance
from NGC 2244.)

Figure 2 shows distributions of the flux and size of clumps in the RMC, as 
a whole and when separated into starless or protostellar clumps.  Figure 2a 
shows histograms of 350 $\mu$m integrated fluxes, $S_{350}$, of the clumps,
demonstrating how the brightest are indeed protostellar (see Figure 1).  
Starless and protostellar clumps with $S_{350}$ $\approx$ 5 Jy are similar 
in number, but clumps fainter than 5 Jy are predominantly starless.  For 
reference (see \S 4 below), a clump of 5 Jy flux at 350 $\mu$m in the RMC 
has a mass of $\sim$7~\msun, assuming $T_{dust}$ = 20 K.  Figure 2b shows 
size histograms of each population where size is the geometric mean of the 
observed major and minor axes of the 2D Gaussian fit to each clump extent.  
(These sizes are not deconvolved from the finite beam.)  Clumps of each type 
have sizes that extend from the resolution limit up to $\sim$0.65 pc.  The 
few larger clumps (i.e., 0.7-1.0 pc) are all protostellar, however.  

\begin{figure}[ht]
\includegraphics[angle=0,width=90mm]{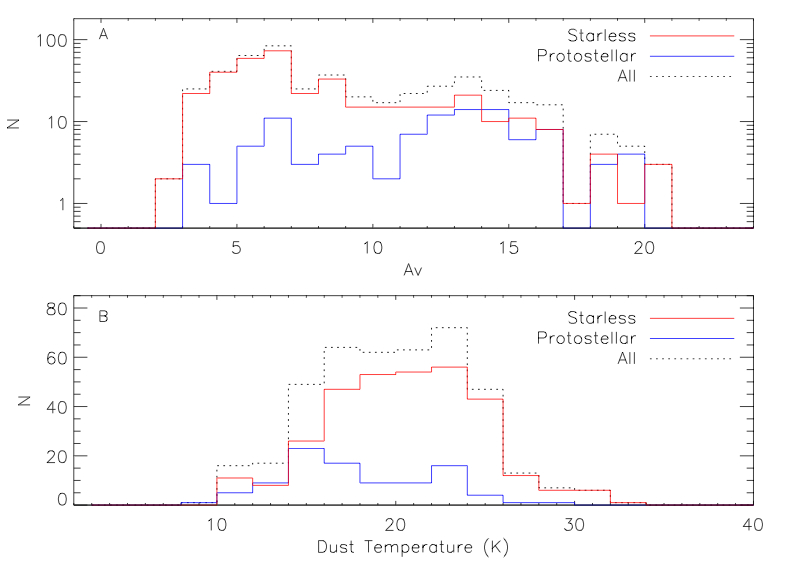}
\caption [] {Distributions of local visual extinctions (A) and temperatures 
(B) for the RMC clumps.  Lines defined as for Figure 2.}
\label{clumptemp}
\end{figure}

Figure 3 shows distributions of the local extinction and temperatures of 
clumps within the cloud, again as a whole and when separated into starless
or protostellar clumps.  Figure 3a shows the $A_{V}$ values associated 
with the clumps, as determined from the low-resolution (i.e., $\sim$2\am) 
extinction maps.  For a given clump, its associated low-resolution $A_{V}$ 
can be used as a proxy for its depth within its parent cloud (see Johnstone, 
Di Francesco \& Kirk 2004).  Clumps of each type are found across the same 
range of extinction ($A_{V}$ $\approx$ 3-20).  No clumps of either type are 
found at $A_{V} <$ 2, however.  Starless clumps are found in greater numbers at 
lower extinctions ($A_{V} <$ 12), but both types are found in similar numbers 
at higher extinctions ($A_{V} \geq$ 12).  The number of starless clumps peaks 
at $A_{V}$ $\approx$ 6, while that of protostellar clumps peaks at $A_{V}$ 
$\approx$ 13-14.  Though starless and protostellar clumps are found over 
the same range of $A_{V}$, the difference in peak $A_{V}$ may be due in part 
to protostellar clumps having higher masses on average than starless clumps 
(see \S 4 below).  Assuming similar sizes, the resulting relatively higher 
column densities could mean relatively higher local $A_{V}$ values, though
the difference in the AvMAP and {\it Herschel} resolutions is large.

Figure 3b shows the distributions of the clump temperatures ($T_{dust}$) found 
across the RMC using all five {\it Herschel} wavelengths and a method described 
by Schneider et al. (2010b)  Temperatures could be found for only 402 of the 
473 identified clumps (i.e., 323 starless and 79 protostellar clumps), and we 
limit further discussion only to those clumps.  Dust temperatures for these 
clumps range between 10 K and 30 K.  Interestingly, starless clumps appear to 
be generally warmer than protostellar clumps, a surprise as the latter could be 
expected to be warmer given internal heating by YSOs.  The difference here, 
however, stems from the large population of starless clumps located to the 
northwest of the RMC (see Figure 1).  At this location, the dust is quite 
warm owing to its proximity to the O stars in NGC 2244.

\section{Clump mass spectra} \label{analysis}

The mass for each clump, $M_{C}$, was determined using its observed 350 \um\ 
flux and the relation 
$M_{C}$ $=$ $S_{\nu}d^{2}$/$\kappa_{\nu}B_{\nu}$($T_{dust}$) = 
0.073($S_{\nu}/\rm{Jy}$)($d/\rm{kpc}$)$^{2}$($\kappa_{\nu}/0.07$ cm$^{2}$g$^{-1}$)$^{-1}$[exp(41.1/$T_{dust}$)-1] \msun.
%\begin{equation}
%M_{C} = 0.073 \left(\frac{S_{350}}{\mbox{Jy}}\right) \left(\frac{d}{\mbox{kpc}}\right)^2 \left(\frac{\kappa_{350}}{0.0708\ \mbox{cm$^2$ g$^{-1}$}}\right)^{-1} \times \left[\exp{\left(\frac{41.11}{T}\right)}-1\right] M_{\odot}
%\end{equation}
%\noindent
To account for variations in temperature across the RMC, we assume for each 
clump its respective $T_{dust}$ value when determining its mass, rather than 
averages of the distributions shown in Figure 3b.  We also use $\kappa_{350}$ 
= 0.07 \cmg for a dust opacity, consistent with the dust prescription of 
Ossenkopf \& Henning (1994), and a distance of 1.6 kpc.

Figure 4 shows the distributions of mass for the total, starless, and 
protostellar clump populations, i.e., their clump mass spectra (CMS).  
Masses for all clumps range across $\sim$3.5 orders of magnitude, i.e., 
-0.75 $\leq$ log~($M_{C}$/\msun) $\leq$ 2.50.  Evidence for differences 
in mass between the populations is seen, however, with starless clumps 
having on average lower masses than protostellar clumps.  Moreover, the 
starless population peaks at log~($M_{C}$/\msun) = 0.25, while the 
protostellar population peaks at log~($M_{C}$/\msun) = 1.0.  This 
difference is likely because the starless clumps in the RMC are both 
warmer and have lower submillimetre fluxes on average than the protostellar 
clumps (see Figures 2 and 3).  Note also that there are fewer higher 
mass starless clumps than protostellar clumps, e.g., over the high-mass 
range of log~($M_{C}$/\msun) = 1.75-2.50, only two starless clumps are
found but 11 protostellar clumps.

\begin{figure}[ht]
\includegraphics[angle=0,width=90mm]{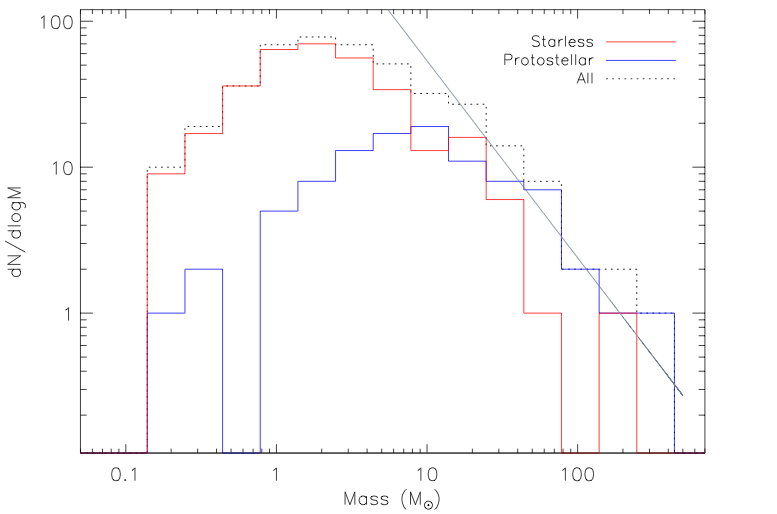}
\caption [] {Clump mass spectra.  Lines defined as for Figure 2.  For 
comparison, the solid black line indicates the slope of the steeper,
high-mass tail of the Salpeter IMF.}
\label{cms}
\end{figure}

Table 1 lists the linear slopes ($\alpha$) fit to the high-mass tails of 
the respective CMS, using the least-squares method and assuming Poissonian 
errors in each mass bin.  Table 1 also lists the mass ranges used to compute
the slopes.  In general, slopes were fit between the bin containing the peak 
of the respective CMS and the bin of the highest mass in the sample.  The 
$\alpha$ = -0.65 $\pm$ 0.01 for all high-mass clumps is slightly more than 
the value of -0.8 found by Dent et al.\/ (2009) from CO 3-2 observations over 
a 5 deg$^{2}$ map of the entire Rosette complex.  The value found here, 
however, is quite similar to the -0.6 slope found by Schneider et al.\/ (1998) 
for higher clump masses using CO (2-1) maps of the Rosette.  Both these groups 
used different clump identification algorithms than used here, i.e., {\it 
clumpfind}\/ by Dent et al.\/ and {\it gaussclumps}\/ by Schneider et al.\/ 
(1998).  Moreover, we have determined masses from clumps we identified here, 
not over the extents of the clumps found by either group using CO.

\begin{table}[ht] 
\label{cmfslopes}
\caption{Slopes of high-mass tails of RMC clump mass spectra}
\centering{
\begin{tabular}{ccc}
\\
\hline
\hline
Clump Type & $\alpha$   &  Mass Range \\ 
\hline
All        & -0.65 $\pm$ 0.01 & 0.25 $\leq$ log~($M_{C}$/\msun) $\leq$ 2.50 \\
Starless   & -0.82 $\pm$ 0.02 & 0.25 $\leq$ log~($M_{C}$/\msun) $\leq$ 2.25 \\
Protostellar & -0.80 $\pm$ 0.02 & 1.00 $\leq$ log~($M_{C}$/\msun) $\geq$ 2.50 \\
\hline
\end{tabular}
}
\end{table}

After dividing the clumps by type, our starless and protostellar clumps have 
similar slopes to their high-mass distributions, i.e., $\alpha$ = -0.8, but 
over different mass ranges.  Indeed, these slopes are more similar to that 
found by Dent et al.  The high-mass tails of the all-clump CMS has a shallower 
slope than do just starless or protostellar clumps.  This difference stems 
from the difference in the mass ranges used to measure slopes.  Also, the 
numbers of protostellar clumps peak at a higher mass than do those of starless 
clumps, and they dominate the all-clump CMS at even higher masses.

With single fits, all three CMS are shallower at higher masses than seen 
in the high-mass tails of mass spectra of small-scale structures within 
closer clouds.  Indeed, the slopes of such tails are seen to be consistent 
with the -1.35 slope of the Salpeter IMF (e.g., in Aquila; see Andr\'e et 
al.\/ 2010).  For comparison with the RMC, a Salpeter slope is also shown 
in Figure 4.  This difference may come simply from resolution; namely, in 
closer clouds, sites of individual star formation, (``cores") can be 
resolved.  For example, typical core separations in Ophiuchus at 125 pc were 
found to be $\sim$0.03 pc by Motte, Andr\'e \& Neri (1998).  Though our minimum 
linear resolution (at 250 \um) is $\sim$0.12 pc, or about the expected size 
of isolated cores, we may not be able to resolve individual cores in the RMC 
owing to crowding on scales $<$ 0.12 pc.  Indeed, such crowding may cause the 
resulting mass spectra to be artificially shallow.  Higher resolution data
will be critical for testing this possibility.

Without appropriate spectral line data, we cannot estimate the states of
gravitaitonal stability for the clumps in the RMC, and determine which may be 
bound or which may be transient structures.  For example, we cannot presently 
guess which starless clumps will likely engage in future star formation.  It is 
clear from Figure 1, however, that the {\it Herschel} data have revealed that 
star formation is ongoing in the RMC and that we are seeing only a snapshot of 
a very dynamical situation.

\section{Summary} \label{summary}

We have conducted a preliminary analysis of small-scale emission within the
Rosette molecular cloud, as observed by the {\it Herschel} Space Observatory.  
From SPIRE and smoothed PACS images, we identified 473 clumps using {\it 
getsources}, a new algorithm that identifies structure over multiple 
wavelengths and scales.  Using {\it Spitzer}\/ data, we classified these into 
371 starless and 102 protostellar clumps.  The clumps have dust temperatures 
of 10-30 K, with starless clumps surprisingly warmer on average than the 
protostellar clumps, owing to the proximity of many to NGC 2244.  Masses 
were determined for each clump, revealing a range of -0.75 $\leq$ 
log~($M_{C}$/\msun) $\leq$ 2.50.  Linear least-squares fits to the high-mass 
tails of the CMS reveal slopes that are consistent with slopes fit to mass 
spectra of clumps identified through CO in the RMC.  Such slopes are shallower 
than those seen in the high-mass tails of the mass spectra of small-scale 
continuum structures (``cores") in closer clouds, possibly from crowding of 
such objects on scales smaller than probed here.

\begin{acknowledgements}
We thank our referee, Jonathan Williams, for comments that improved this 
paper.  JDF acknowledges support by the National Research Council of Canada, 
the Canadian Space Agency (via a SSEP Grant), and the Natural Sciences and 
Engineering Council of Canada (via a Discovery Grant).  JDF also thanks David
Naylor for encouragement and assistance over the years.\\
SPIRE has been developed by a consortium of institutes led by Cardiff 
University (UK) and including Univ. Lethbridge (Canada); NAOC (China); CEA, 
LAM (France); IFSI, Univ. Padua (Italy); IAC (Spain); Stockholm Observatory
(Sweden); Imperial College London, RAL, UCL-MSSL, UKATC, Univ. Sussex (UK); 
and Caltech, JPL, NHSC, Univ. Colorado (USA). This development has been 
supported by national funding agencies: CSA (Canada); NAOC (China); CEA, 
CNES, CNRS (France); ASI (Italy); MCINN (Spain); SNSB (Sweden); STFC (UK); 
and NASA (USA). PACS has been developed by a consortium of institutes led 
by MPE (Germany) and including UVIE (Austria); KUL, CSL, IMEC (Belgium); 
CEA, LAM (France); MPIA (Germany); IFSI, OAP/AOT, OAA/CAISMI, LENS, SISSA 
(Italy); IAC (Spain). This development has been supported by the funding 
agencies BMVIT (Austria), ESA-PRODEX (Belgium), CEA/CNES (France), DLR 
(Germany), ASI (Italy), and CICT/MCT (Spain).
\end{acknowledgements}


\begin{thebibliography}{}

\bibitem[2010]{andre2010}
Andr\'e, P. et al. 2010, A\&A, this volume

\bibitem[2009]{dent2009}
Dent, W. R. F., et al.\/ 2009, MNRAS, 395, 1805

\bibitem[2010]{griffin2010}
Griffin, M., et al. 2010, this volume

%Henning, T., Michel, B., \& Stognienko, R., 1995, Planet. Space Sci., 43, 1333
%
%\bibitem[19XX]{hildebrand19XX}
%Hildebrand, R. et al. 19XX, ApJ, XXX, YYY

\bibitem[2010]{henneman2010}
Henneman, M., et al. 2010, this volume

\bibitem[1962]{johnson1962}
Johnson, H. L. 1962, ApJ, 136, 1135

\bibitem[2004]{jdk2004}
Johnstone, D., Di Francesco, J. \& Kirk, H. 2004, ApJ, 611, L45

\bibitem[2005]{ls2005}
Li, J. Z., \& Smith, M. 2005, AJ, 130, 721

%\bibitem[1998]{lis1998}
%Lis, D., Serabyn, E., Keene, J., Dowell, C. D., Benford, D. J., Phillips, T.
%G., Hunter, T. R., \& Wang, N. 1998, ApJ, 299, 308

\bibitem[2001]{la2001}
Lombardi, M., \& Alves, J. 2001, A\&A, 377, 1023

\bibitem[2010]{sasha2010}
Men'shchikov, A. et al. 2010, A\&A, this volume

\bibitem[1998]{man1998}
Motte, F., Andr\'e, P., \& Neri, R. 1998, A\&A, 336, 150

\bibitem[2010]{motte2010}
Motte, F., et al. 2010, A\&A, this volume

\bibitem[1994]{oh1994}
Ossenkopf, V., \& Henning, T. 1994, A\&A, 291, 943

\bibitem[2010]{ott2010}
Ott, S. 2010, in ASP Conference Series, Astronomical Data Analysis Software 
and Systems XIX, Y. Mizumoto, K.-I. Morita, and M. Ohishi, eds. in press

\bibitem[2001]{ps2002}
Park, B., \& Sung, H. 2002, AJ, 123, 892

\bibitem[1987]{perez1987}
P\'erez, M. R., Th\'e, P.-S., \& Westerlune, B. E. 1987, PASP, 99, 1050

\bibitem[1998]{pl1998}
Phelps, R., \& Lada, E. A. 1998, ApJ, 477, 176

\bibitem[2010]{pilbratt1010}
Pilbratt, G., et al. 2010, this volume

\bibitem[2010]{poglitsch1010}
Poglitsch, A., et al. 2010, this volume

\bibitem[2008]{poulton2008}
Poulton, C. J., Robitaille, T. P., Greaves, J. S., Bonnell, I. A., Williams, 
J. P., \& Heyer, M. H. 2008, MNRAS, 384, 1249

\bibitem[2008]{roman2008}
Rom\'an-Z\'u\~niga, C. G., Elston, R., Ferreira, B., \& Lada, E. A. 2008, 
ApJ, 672, 861

\bibitem[2008]{rzl2008}
Rom\'an-Z\'u\~niga, C. G., \& Lada, E. A. 2008, in ``The Handbook of Star 
Forming Regions, Volume 1: The Northern Hemisphere," ed. B. Reipurth, (ASP: 
San Francisco), p. 928 (RZL)

%\bibitem[2010]{sadavoy2010}
%Sadavoy, S., Di Francesco, J., Bontemps, S., Megeath, S. T., Rebull, L. M.,
%Allgaier, E., Carey, S., Gutermuth, R., Hora, J., Huard, T., McCabe, C.-E.,
%Muzerolle, J., Noreiga-Crespo, A., Padgett, D. \& Terebey, S. 2010, ApJ, 
%710, 1247

\bibitem[2010]{sadavoy2010}
Sadavoy, S. et al.\/ 2010, ApJ, 710, 1247

\bibitem[2010]{sch2010a}
Schneider, N., Bontemps, S., Simon, R., et al. 2010a, astro-ph/1001.2453

\bibitem[2010]{sch2010b}
Schneider, N. et al.\/ 2010b, this volume

\bibitem[1998]{schenider1998}
Schneider, N., Stutzki, J., Winnewisser, G., \& Block, D. 1998, A\&A, 335, 1045

%\bibitem[2010]{sb2010}
%Swift, J. J., \& Beaumont, C. N. 2010, PASP, 122, 224

\bibitem[2000]{wbm2000}
Williams, J. P., Blitz, L., \& McKee, C. F. 2000, in ``Protostars and Planets
IV," eds. V. Mannings, A. P. Boss, \& S. S. Russell, (Tucson: University of 
Arizona), p. 97

\bibitem[1995]{wbs1995}
Williams, J. P., Blitz, L., \& Stark, A. 1995, ApJ, 451, 252

\end{thebibliography}
\end{document}